\documentclass[]{pasj01}
\usepackage[switch,mathlines]{lineno}

\Received{2024 July 18}%{yyyy/mm/dd}
\Accepted{2024 September 18}%{yyyy/mm/dd}
\jyear{2024}

\newcommand{\Ni}{\ensuremath{^{56}\mathrm{Ni}}\xspace}
\newcommand{\Co}{\ensuremath{^{56}\mathrm{Co}}\xspace}

\newcommand{\Msun}{\,\ensuremath{\mathrm{M}_\odot}\xspace}
\newcommand{\Rsun}{\,\ensuremath{\mathrm{R}_\odot}\xspace}
\newcommand{\Msunpyr}{\,\ensuremath{\Msun~\mathrm{yr^{-1}}}\xspace}
\newcommand{\kmps}{\,\ensuremath{\mathrm{km~s^{-1}}}\xspace}
\newcommand{\Mni}{\ensuremath{\mathrm{M}_{^{56}\mathrm{Ni}}}\xspace}

\graphicspath{{./}{figures/}}

\DeclareAbbreviation\pasa{PASA}

\usepackage{xspace}
\begin{document} 

\title{ 
\LETTERLABEL %%% <-- uncomment for LETTER article  
%\REVIEWLABEL %%% <-- uncomment for REVIEW article  
Blue supergiants as a progenitor of intermediate-luminosity red transients
}

%%% begin:list of authors
% Do NOT capitalize all letters in "textsc".
\author{Takashi J. \textsc{Moriya}\altaffilmark{1,2,3}}
\altaffiltext{1}{National Astronomical Observatory of Japan, National Institutes of Natural Sciences, 2-21-1 Osawa, Mitaka, Tokyo 181-8588, Japan}
\altaffiltext{2}{Graduate Institute for Advanced Studies, SOKENDAI, 2-21-1 Osawa, Mitaka, Tokyo 181-8588, Japan}
\altaffiltext{3}{School of Physics and Astronomy, Monash University, Clayton, VIC 3800, Australia}
\email{takashi.moriya@nao.ac.jp}

\author{Athira \textsc{Menon}\altaffilmark{4,5,6}}
\altaffiltext{4}{Instituto de Astrofísica de Canarias, Avenida Vía Láctea s/n, 38205 La Laguna, Tenerife, Spain}
\altaffiltext{5}{Universidad de La Laguna, Departamento de Astrofísica, Avenida Astrofísico Francisco Sánchez s/n, 38206 La Laguna, Tenerife, Spain}
\altaffiltext{6}{Columbia University, Department of Astronomy, New York, NY 10027, USA}
\email{aam2371@columbia.edu}

%% `\KeyWords{}' always has to be placed before ``\maketitle'' 
%%  List of Key Words:  https://academic.oup.com/pasj/pages/Pasj_Keywords 
\KeyWords{supernovae: general --- stars: massive --- binaries: close}  

\maketitle

\begin{abstract}
The current perspective about the explosions of massive hydrogen-rich blue supergiants is that they resemble SN~1987A. These so-called peculiar Type~II supernovae, however, are one of the rarest types of supernovae and may not hence be the fate of all blue supergiants. In this work, we explore other explosion scenarios for blue supergiants. We create synthetic light curves from the explosions of blue supergiant models born from binary mergers, over a range of explosion energies and \Ni masses. We find that blue supergiant explosions may also lead to intermediate-luminosity red transients. We thus identify two categories of supernovae possible from blue supergiant explosions: those with high \Ni masses ($\gtrsim 0.01$\Msun) result in slow-rising, dome-shaped light curves like SN~1987A. Lower \Ni masses result in low-luminosity, short-plateau light curves resembling some intermediate-luminosity red transients and Type~II supernovae like SN~2008bp, which are possible from the explosions of compact blue supergiants and not from the far more extended red supergiants. Our results indicate that blue supergiant explosions are more diverse than SN~1987A-like events and may be hidden among different kinds of transients, explaining the possible discrepancies between the expected fraction of blue supergiants born from binary mergers and the observed fraction of SN~1987A-like supernovae.
\end{abstract}

\clearpage

%\pagewiselinenumbers

\begin{longtable}{cccccccccc}
  \caption{Properties of BSG progenitor models at core-carbon depletion used in this work, selected from \citet{menon2024}.}\label{tab:models}
  \hline              
$M_\textrm{1,i}^{(a)}$ & $M_\textrm{2,i}^{(b)}$ & $P_\textrm{i}^{(c)}$ & $f_\textrm{c}^{(d)}$  &  log\,$L_{\star}^{(e)}$ & T$_\textrm{eff,$\star$}^{(f)}$ & $M_{\star}^{(g)}$ & $M_{\star}^{\textrm{env} (h)}$ & $M_{\star}^{\textrm{core} (i)}$ & $R_{\star}^{(j)}$ \\
(\Msun) & (\Msun) & (d) & & (L$_{\odot}$) & (kK) & (\Msun) & (\Msun) & (\Msun) & (\Rsun)\\
\endfirsthead
  \hline
\endhead
  \hline
\endfoot
  \hline
\multicolumn{10}{l}{$^{(a)}$ Initial mass of the primary of the pre-merger binary.}\\
\multicolumn{10}{l}{$^{(b)}$ Initial mass of the secondary of the pre-merger binary.}\\  
\multicolumn{10}{l}{$^{(c)}$ Initial orbital period of the pre-merger binary.}\\  
\multicolumn{10}{l}{$^{(d)}$ Core dredge-up factor at the time of the merger as defined in \citet{menon2024}.}\\
\multicolumn{10}{l}{$^{(e)}$ Luminosity of post-merger star at central-carbon depletion.}\\
\multicolumn{10}{l}{$^{(f)}$ Effective temperature of post-merger star at central-carbon depletion.} \\
\multicolumn{10}{l}{$^{(g)}$ Stellar mass of post-merger star at central-carbon depletion.} \\
\multicolumn{9}{l}{$^{(h)}$ Hydrogen-rich envelope mass of post-merger star at central-carbon depletion.} \\
\multicolumn{9}{l}{$^{(i)}$ Hydrogen-free core mass of post-merger star at central-carbon depletion.} \\
\multicolumn{9}{l}{$^{(j)}$ Stellar radius of post-merger star at central-carbon depletion.} \\
\endlastfoot
  \hline
11.2 & 9.0 & 1413 & 0.06 & 5.0 & 30.6 & 16.9 & 13.2 & 3.7 & 10.8 \\
20.0 & 15.0 & 3162 & 0.08 & 5.6 & 33.8 & 32.3 & 25.7 & 6.6 & 17.8 \\
20.0 & 2.0 & 16 & 0.00 & 5.2 & 18.8 & 19.7 & 12.6 & 7.1 & 37.4 \\
31.6 & 13.3 & 28 & 0.28 & 5.5 & 20.3 & 28.6 & 17.0 & 11.6 & 50.8 \\
25.1 & 1.3 & 1000 & 0.00 & 5.4 & 15.0 & 24.6 & 14.7 & 9.9 &  76.0 \\
%39.8 & 20.7 & 631 & 0.14 & 5.9 & 17.5 & 40 & & 96.0 \\
\end{longtable}

\section{Introduction}
Blue supergiants (BSGs), with spectral classes between B0 and BI, are observed in plenty in the local and distant Universe \citep{menon2024}. Yet, despite their abundance, the nature of their explosive fates has not been thoroughly explored. SN~1987A was the explosion of a BSG, whose progenitor was imaged to be a BIa supergiant \citep{sonneborn1987}. Indeed, the progenitor properties of SN~1987A can be well explained via stellar-merger scenario \citep{podsiadlowski1992,menon2017,urushibata2018}, including the features of the supernova (SN) light curve as well, along with a few other SNe with similar light-curve shapes \citep{menon2019}. Collectively these SN~1987A-like SNe are called ``peculiar Type~II SNe.''

Recently, \citet{menon2024} demonstrated that the majority of BSGs observed today have been born from binary mergers of post-main sequence Hertzsprung-gap or red supergiants (RSGs) with their main-sequence companions. They further predicted that a substantial fraction of the observed BSGs may also die as BSGs. Stellar mergers are expected to be the outcome of  about $\sim30\%$ of massive binary star systems \citep{demink2014}, of which the majority may occur during the post-main sequence expansion of the primary star. SN~1987A-like events, however, are rare, comprising only $\lesssim5$\% of the known core-collapse SNe (e.g., \cite{kleiser2011,pastorello2012,taddia2016,sit2023}). This fraction is far fewer than the observed fraction of BSGs among massive star populations, a large number of which are expected to remain as BSGs when they undergo core-collapse and explode, according to the binary-merger models of \citet{menon2024}.

The light curve of SN~1987A has a slow rise lasting for about 80~days \citep{arnett1989}, while typical Type~II SNe have a luminosity plateau lasting for around 100~days \citep{anderson2014}. The explosions of BSGs have been mostly associated with the characteristic slow-rising, dome-shaped light curves of SN~1987A which are due to the compactness of BSGs (R$_{\star}\lesssim 70$\Rsun, e.g., \cite{taddia2016,sit2023} but see also \cite{kleiser2011,tsuna2020}). Due to the small radii of BSGs, the adiabatic expansion after the explosion is efficient and, therefore, the luminosity declines quickly at first. Then, the SN ejecta are heated by the nuclear decay energy of \Ni and this \Ni heating makes the characteristic slow rise of SN~1987A (e.g., \cite{woosley1988,shigeyama1990,blinnikov2000,dessart2019,pumo2023}). In the case of SN~1987A, the amount of \Ni\ synthesized at the explosion is estimated to be around 0.07~\Msun \citep{arnett1989}. 

Since the slow light-curve rise observed in SN 1987A-like SNe is powered by \Ni decay heating, this characteristic slow light-curve rise would become less significant if the amount of \Ni synthesized during the explosions of BSGs is small. In such a case, BSG explosions might not be observed as SN~1987A-like events and it is possible that a population of BSG explosions do not become SN~1987A-like events. 

Intermediate-luminosity red transients (ILRTs) are a class of ``gap transients'' which have peak absolute magnitudes between those of classical novae and SNe (\cite{rau2009,kasliwal2012}; see \cite{pastorello2019} for a review). They are characterized by a low-luminosity plateau lasting for a couple of months (e.g., \cite{cai2021}). The late-time light-curve decline rate is consistent with those of the \Co decay rate from $\Mni\sim 10^{-3}-10^{-4}~\Msun$, indicating that they could be related to terminal SN explosions (e.g., \cite{cai2021}). Because of their faint nature as well as the low-mass progenitor observed for SN~2008S \citep{prieto2008,adams2016}, ILRTs have been related to low mass core-collapse SNe such as electron-capture SNe (e.g., \cite{botticella2009,cai2021}, but see also \cite{tominaga2013,moriya2014,kozyreva2021} where electron-capture SNe are suggested to become more luminous than ILRTs). In the case of a potential ILRT AT~2019krl, a BSG was identified as a progenitor \citep{andrews2021}. Thus ILRTs can come from several different progenitors.

In this work, we explore the light-curve properties of BSG explosions over a range of explosion energies ($E=0.1-1$~B, where $1~\mathrm{B}\equiv10^{51}~\mathrm{erg}$) and \Ni masses ($\Mni=0.001-0.07\Msun$). The progenitor models are taken from \citet{menon2024} which are evolved from detailed stellar-merger models until core carbon-depletion and have masses of $16-40$\Msun. We identify a dichotomy in SNe from BSG explosions, which depends mainly on the \Ni mass (\Mni) ejected: for $\Mni\gtrsim 0.01~\Msun$ we get SN~1987A-like light curves and for $\Mni\lesssim 0.01~\Msun$ we get low-luminosity, short-plateau light curves that are consistent with certain ILRTs and Type~II SNe like SN~2008bp.

The rest of this paper is organized as follows. We first introduce the BSG progenitor models and our methods of SN light-curve calculations in Section~\ref{sec:method}. We present the results of the light-curve calculations in Section~\ref{sec:results} and show that BSG explosions can result in low-luminosity, short-plateau Type~II SNe when the amount of \Ni is sufficiently small. We compare the low-luminosity, short-plateau light curves from BSGs with those of ILRTs in Section~\ref{sec:ilrts} and show that they are indeed similar. We summarize our discoveries of this paper in Section~\ref{sec:summary}.

\begin{figure}
 \begin{center}
  \includegraphics[width=0.9\columnwidth]{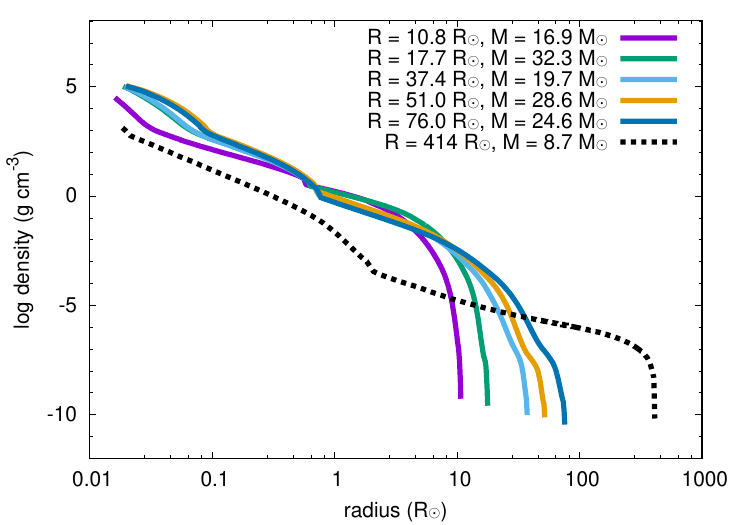}
  \includegraphics[width=0.9\columnwidth]{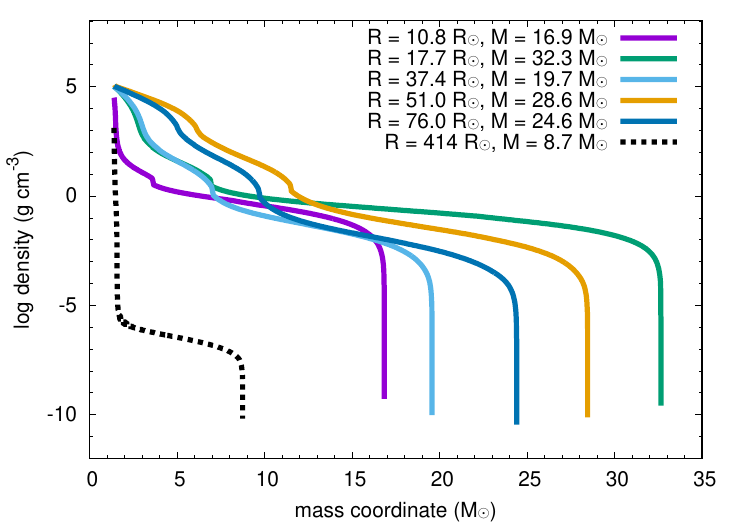}  
 \end{center}
\caption{
Density structure of BSG progenitors adopted in this work. The top panel shows the structure in the radius coordinate and the bottom panel shows the structure in mass coordinate. The structure of a RSG progenitor model from \citet{sukhbold2016} is presented with the dashed line for comparison. The RSG model is evolved until core collapse, while the BSG models are evolved until core carbon depletion.
}\label{fig:progenitor_structure}
\end{figure}

\section{Methods}\label{sec:method}
\subsection{Progenitor models}

The BSG models used in this paper are from \citet{menon2024}, which have been evolved through a binary merger of a core hydrogen-depleted giant (on the Hertzsprung gap or as a RSG) with its main-sequence companion. These post-merger models evolved to core He-burning BSGs and matched the spectroscopic properties of the majority of the sample of early B-type supergiants in the Large Magellanic Cloud (LMC). Those models with masses $\approx16-25$\,M$_{\odot}$ always exploded as BSGs, while those with higher masses (upto 40\,M$_{\odot}$) either exploded as BSGs or as cooler stars with T$_\textrm{eff}\leq10$\,kK, depending on the dynamics of the merging process.

The outcomes of the merger models primarily depend on the core-envelope structure and the power of the shell burning regions, with winds only playing a minor role. As the LMC metallicity is only half the solar value, we do not expect winds to make a significant contribution to our models, within the mass range considered. Hence our LMC merger models can be applied to study Galactic BSGs as well.

For this study, we select representative BSG progenitor models that span the upper and lower mass range of B-type supergiants. These models are evolved until core carbon-depletion and summarized in Table~\ref{tab:models}. After the core carbon depletion, the envelope structure determining light-curve properties is not expected to change much until core collapse. The density structure of the BSG models are presented in Fig.~\ref{fig:progenitor_structure}. In the same figure, we also show the 8.7~\Msun (9.0~\Msun at the zero-age main sequence [ZAMS]) RSG SN progenitor model (414~\Rsun) from \citet{sukhbold2016} for comparison.

\subsection{Explosion and light-curve modeling}
We conduct explosion and subsequent light-curve modeling by using one-dimensional radiation hydrodynamics code \texttt{STELLA} \citep{blinnikov1998,blinnikov2000,blinnikov2006}. The central 1.4~\Msun of the progenitor models is removed as a compact remnant and the remaining structure is put as the initial condition for the \texttt{STELLA} calculations. The explosion is triggered by inserting thermal energy at the innermost layers of the progenitors above the mass cut at 1.4~\Msun. \texttt{STELLA} does not treat explosive nucleosynthesis and the amount of $^{56}$Ni is a free parameter. In this work, $^{56}$Ni is uniformly mixed within the entire ejecta. The energy input from the gamma-rays from the nuclear decay is treated in a one-group radiation transport approximation \citep{swartz1995}. The positrons are deposited in situ.

Evolution of spectral energy distributions (SEDs) is numerically followed in each timestep in \texttt{STELLA}. We adopt 100 wavelength bins between 1~\AA\ and 50,000~\AA\ in a log scale. We can obtain light curves in any photometic bands by convolving the SEDs with any filter response functions. We present bolometric light curves as well as the \textit{V} and \textit{R}-band light curves in the Johnson-Cousins filter system \citep{bessell1990} in this paper.

\begin{figure}
 \begin{center}
  \includegraphics[width=0.9\columnwidth]{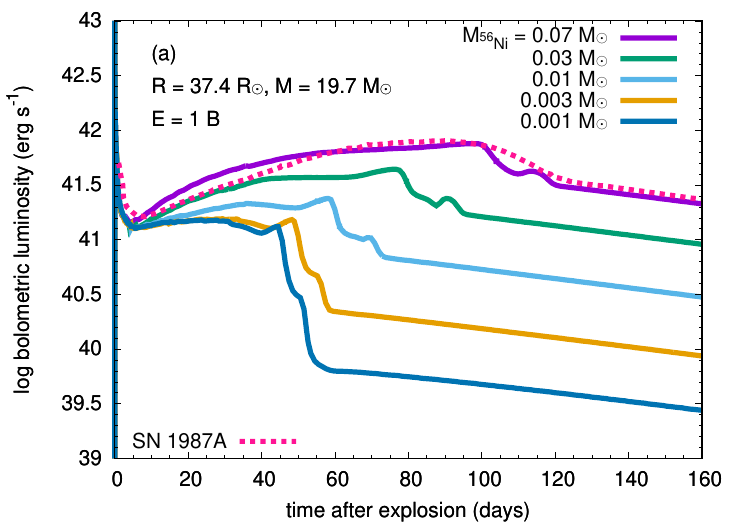}
  \includegraphics[width=0.9\columnwidth]{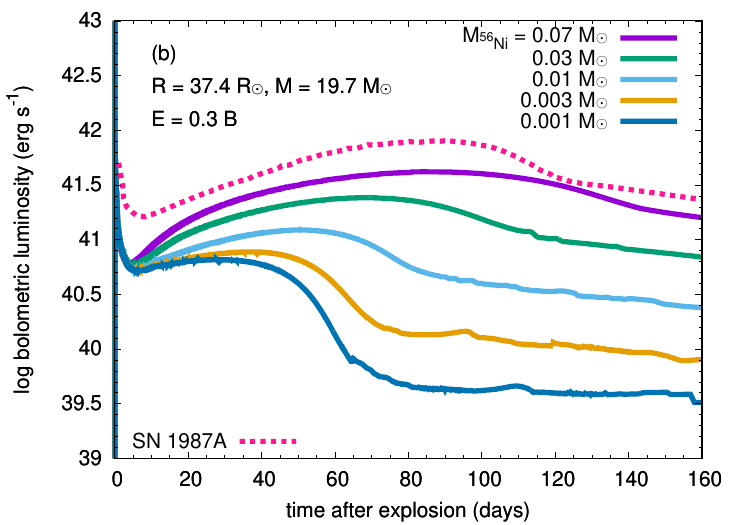}
 \end{center}
\caption{
Effects of \Ni mass in shaping bolometric light curves of BSG explosions. The top panel shows the $1~\mathrm{B}$ models and the bottom panel shows the $0.3~\mathrm{B}$ models. We fix the progenitor model ($R=37.4~\mathrm{R_\odot}$ and $M=19.7~\mathrm{M_\odot}$) in this figure. When the \Ni mass is large enough, we find the luminosity peak powered by the \Ni heating becomes apparent and the SN~1987A-like SN light curves can be reproduced. When the \Ni mass becomes too low, we stop finding the light curve peak powered by the \Ni heating and BSG explosions end up to have low-luminosity, short-plateau light curves. The bolometric light curve of SN~1987A \citep{hamuy1988} is presented as a reference.
}\label{fig:demo}
\end{figure}

\begin{figure*}
 \begin{center}
  \includegraphics[width=0.9\columnwidth]{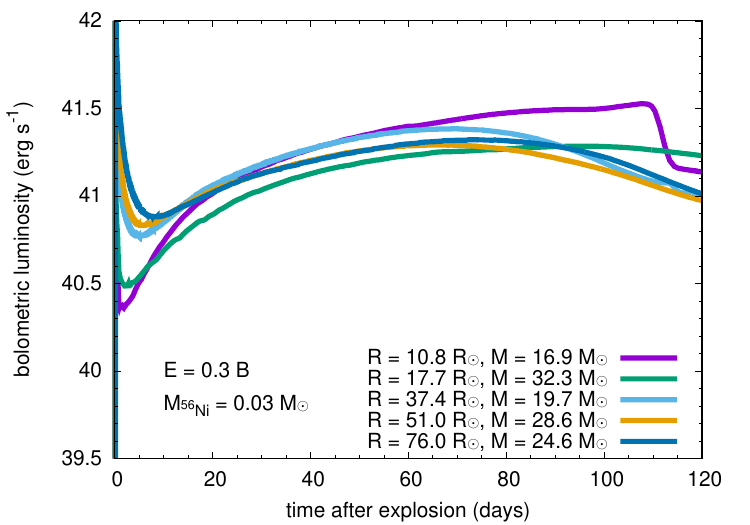}
  \includegraphics[width=0.9\columnwidth]{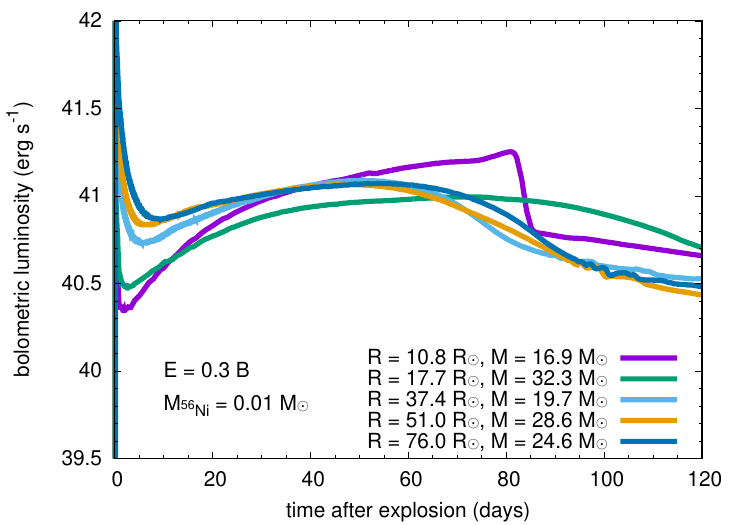} \\
  \includegraphics[width=0.9\columnwidth]{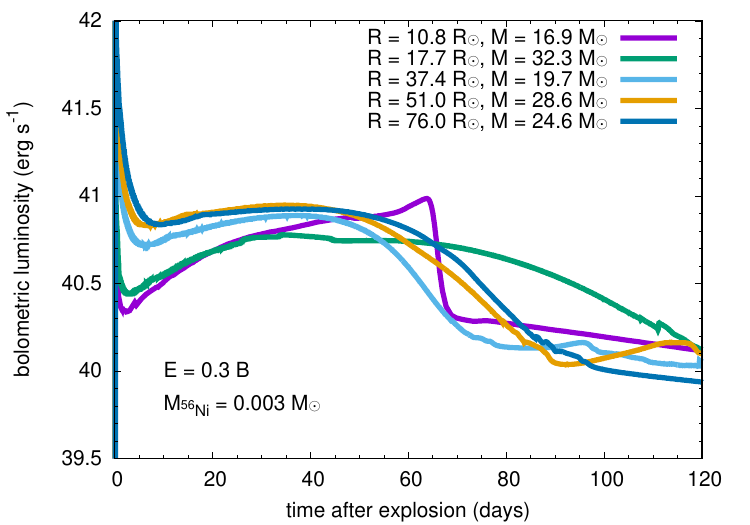} 
  \includegraphics[width=0.9\columnwidth]{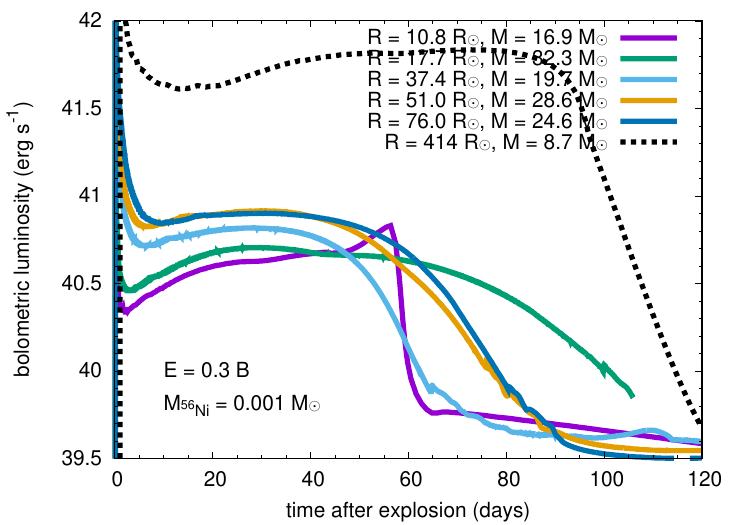}
 \end{center}
\caption{
Bolometric light curves from the explosions of BSGs with various radii ($10.8-76.0~\Rsun$) with $\Mni = 0.03, 0.01, 0.003$ and $0.001~\Msun$ and $E=0.3~\mathrm{B}$. A light curve of a RSG explosion ($R=414~\Rsun$, $M=8.7~\Msun$ with the ZAMS mass of 9.0~\Msun in \cite{sukhbold2016}) with $\Mni=0.001~\Msun$ is presented for comparison (dashed line). In one case, the numerical computation was terminated before reaching 120~days because of numerical issues.
}\label{fig:bol}
\end{figure*}

\section{Results}\label{sec:results}
Fig.~\ref{fig:demo} presents how \Ni\ mass affects the light curves of BSG explosions based on a progenitor model for SN~1987A. This progenitor has the radius $R=37.4~\Rsun$ and the mass $M=19.7~\Msun$. This mass does not exclude the compact remnant mass of 1.4~\Msun and the ejecta mass in the calculation is 18.3~\Msun. In Fig.~\ref{fig:demo}a, we assume the explosion energy $E=1~\mathrm{B}$. We can find that the bolometric light curve of the explosion of this progenitor model with $\Mni=0.07~\Msun$ matches well to that of SN~1987A. 

As we decrease \Mni from this model, we can find that the light-curve peak caused by the \Ni heating become earlier and fainter. Then, when the \Ni mass becomes sufficiently low ($\Mni\lesssim 0.003~\Msun$ in the example in Fig.~\ref{fig:demo}a), we no longer find the luminosity increase due to the \Ni heating. The previous study by \citet{pumo2023} also found that the effects of the \Ni heating become negligible with a similar \Ni mass. When the effects of the \Ni heating become negligible, the luminosity is mainly powered by the thermal energy provided by the initial shock wave triggered the explosion as in the case of Type~II SNe from more extended RSGs. Because BSGs are compact, the adiabatic cooling becomes efficient and the plateau luminosity and duration become low. In the example presented in Fig.~\ref{fig:demo}, we can find that the plateau luminosity is around $10^{41}~\mathrm{erg}$ with the plateau duration of around 50~days. Such a short, low-luminosity plateau is difficult to be realized by RSG explosions because low explosion energy required to realize a faint plateau luminosity makes a plateau duration even longer (e.g., \cite{kasen2009}).

Because \Mni is required to be small not to have a luminosity peak powered by the \Ni heating, it would be natural to investigate BSG explosions with a small explosion energy. In Fig.~\ref{fig:demo}b, we present the bolometric light curves of BSG explosions with $E=0.3~\mathrm{B}$ with various \Ni masses. With the lower explosion energy, we still find that the effect of the \Ni\ heating disappears with $\Mni\lesssim 0.003~\Msun$. The pleateau luminosity decreases and the plateau duration increases with the lower explosion energy in a similar way as in Type~II SNe from RSGs.

Fig.~\ref{fig:bol} explores the effects of the \Ni\ mass on the light curves of the explosions of BSGs with various radii. The adiabatic cooling depends on the progenitor radii, and they are expected to affect the properties of BSG explosions with low \Mni. All the BSG explosion models present the \Ni-powered slow rise in their light curves when $\Mni = 0.03~\Msun$. When $\Mni = 0.01~\Msun$, the slow rise from the \Ni heating starts to be insignificant in the models with $R=37.4~\Rsun$, 51.0~\Rsun, and 76.0~\Rsun, and they only show the plateau phase when $\Mni = 0.003~\Msun$. The $R=17.7~\Rsun$ BSG explosion shows the low-luminosity, short plateau with insignificant luminosity increase with $\Mni= 0.001~\Msun$. Our most compact BSG model ($R=10.8~\Rsun$) may still have a small effect of the \Ni heating with $\Mni\simeq 0.001~\Msun$. We can also find that low-luminosity, short-plateau light curves are difficult to be realized by RSG explosions. Overall, BSG explosions need to have $\Mni \lesssim 0.01~\Msun$ in order not to have a luminosity peak so that they can be observed as low-luminosity, short-plateau Type~II SNe. This critical mass is smaller than the average \Ni mass estimated in Type~II SNe ($\simeq 0.04~\Msun$, \cite{anderson2019}).

\begin{figure}[t]
 \begin{center}
  \includegraphics[width=0.9\columnwidth]{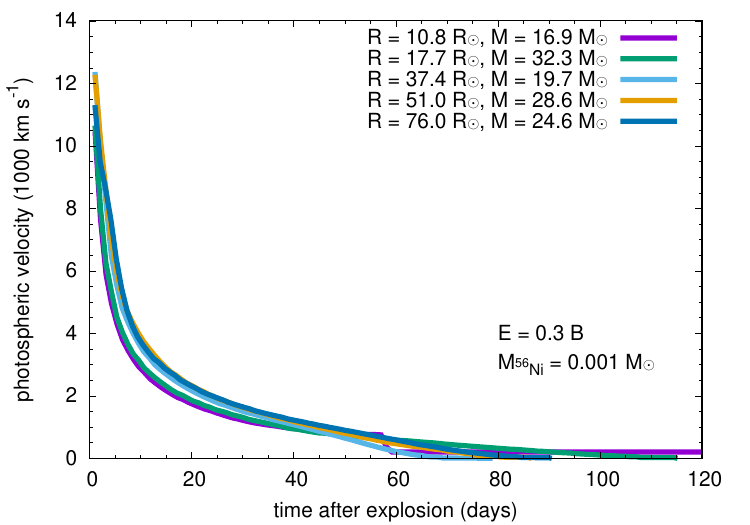}
 \end{center}
\caption{
Photospheric velocity evolution of the BSG explosion models with $\Mni = 0.001~\Msun$ and $E=0.3~\mathrm{B}$. Photosphere is set where the Rosseland-mean optical depth becomes $2/3$.
}\label{fig:photo_vel}
\end{figure}

Because the plateau phase is caused by the recombination within the hydrogen-rich envelope as in the case of RSG explosions, the photospheric temperature during the plateau phase is found to be around 5000~K. The photospheric velocities can be small when the explosion energy is small as presented in Fig.~\ref{fig:photo_vel}.

\begin{figure}
 \begin{center}
  \includegraphics[width=0.9\columnwidth]{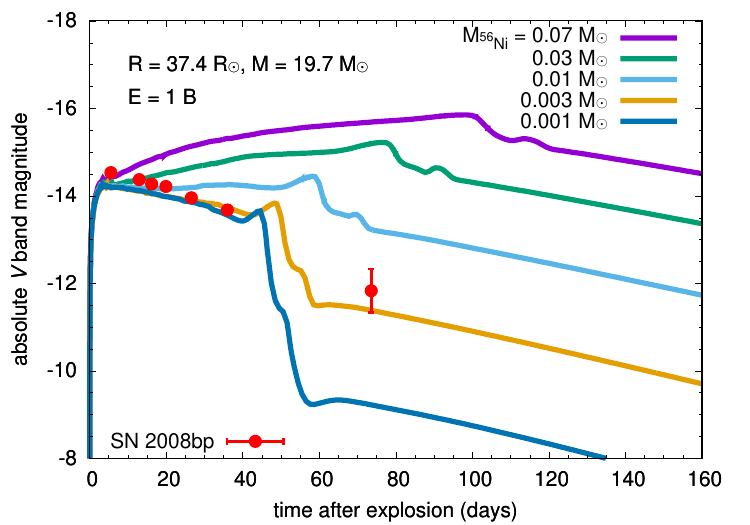}
  \includegraphics[width=0.9\columnwidth]{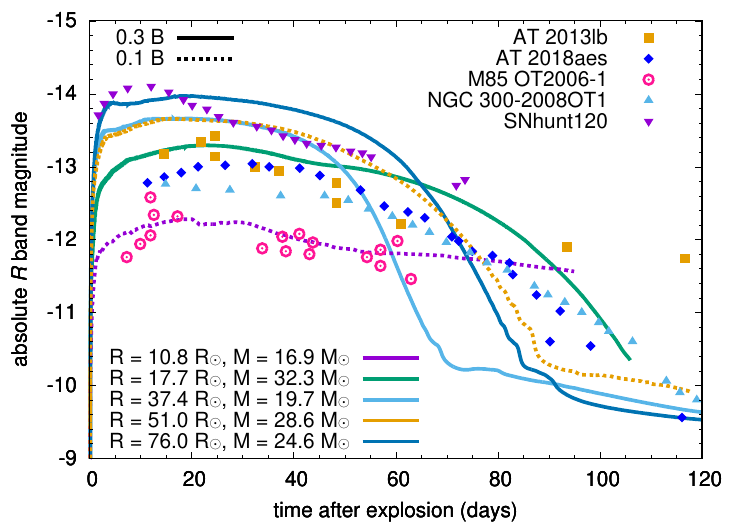}
  \includegraphics[width=0.9\columnwidth]{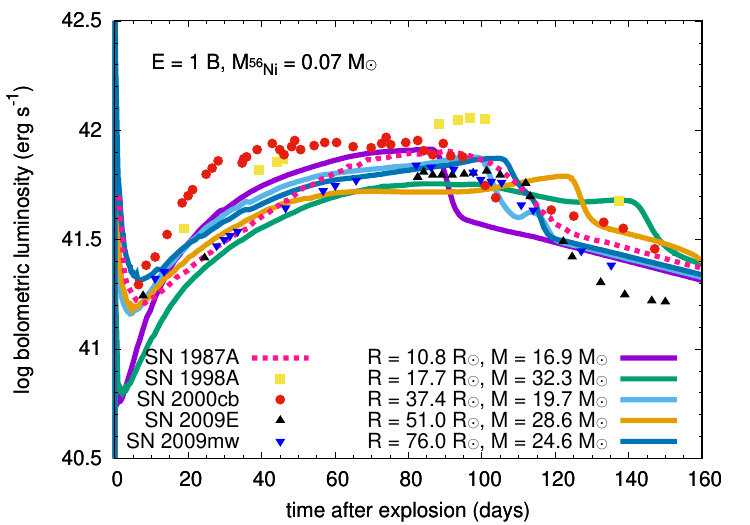}
 \end{center}
\caption{
\textit{Top:} \textit{V}-band light curves of the BSG explosion models presented in Fig.~\ref{fig:demo}a. We can find that the low-luminosity, short-plateau Type~II SN~2008bp reported in \citet{anderson2014} can be explained by the BSG explosion with $\Mni=0.003~\Msun$.
\textit{Middle:} The \textit{R}-band light curves of some BSG explosions with $\Mni=0.001~\Msun$
are compared to those of ILRTs. Their overall light-curve features are similar in many cases. In one case, the numerical computation was terminated before reaching 120~days because of numerical issues. The light curves of AT~2013lb \citep{cai2021}, AT~2018aes \citep{cai2021}, M85~OT2006-1 \citep{kulkarni2007,pastorello2007}, NGC~300-2008OT1 \citep{berger2009,bond2009,prieto2009ot,kashi2010,humphreys2011}, and SNhunt120 \citep{stritzinger2020} are presented.
\textit{Bottom:} The bolometric light curves of BSG explosions with $\mathrm{E=1~B}$ and $\Mni=0.07~\Msun$ compared with those of SN~1987A \citep{hamuy1988} and other SN~1987A-like SNe: SN~1998A \citep{pastorello2005}, SN~2000cb \citep{kleiser2011,utrobin2011}, SN~2009E \citep{pastorello2012}, and SN~2009mw \citep{takats2016}. 
}\label{fig:comparison}
\end{figure}

\section{Comparison with observations}\label{sec:ilrts}
Among Type~II plateau SNe,  low-luminosity ($\sim 10^{41}~\mathrm{erg~s^{-1}}$), short-duration plateaus of $\sim 50~\mathrm{days}$ are quite rare. Still, in the Type~II SN light-curve sample compiled by \citet{anderson2014}, one Type~II SN, SN~2008bp, is found to have such a property and its light curve indeed match well to our BSG explosion model with a small amount of \Ni having no slow rise powered by the \Ni heating (the top panel of Fig.~\ref{fig:comparison}).

The more common variety of low-luminosity, short-plateau light curves are found among the sample of ILRTs.  In the middle panel of Fig.~\ref{fig:comparison}, we compare our low-luminosity, short-plateau BSG explosion models with some representative ILRTs. The low-luminosity plateau phase of ILRTs lasting for a couple of months is consistent with the predicted properties of low-energy, small \Ni mass explosions of BSGs discussed in the previous section. Indeed, their overall light-curve features are found to be similar to ILRTs. Some differences such as the early luminosity bump in, e.g., SNhunt120, may be caused by circumstellar interaction, for example. The low photospheric velocities after the luminosity peak observed in ILRTs (e.g., \cite{cai2021}) are also consistent with our BSG models (Fig.~\ref{fig:photo_vel}). Therefore, we propose that a fraction ILRTs may originate from faint BSG explosions with a small amount of \Ni.

ILRTs are often found to show the signatures of dense circumstellar matter (CSM). While SN~1987A did not show such a signature \citep{chevalier1987}, it is possible that the BSG progenitors from mergers have dense CSM \citep{justham2014}. Assuming the ejecta mass of $30~\Msun$ and the explosion energy of $0.3~\mathrm{B}$, the luminosity from the interaction to the CSM with the mass-loss rate of $10^{-3}~\Msunpyr$ and the wind velocity of $100~\kmps$ is of the order of $10^{40}~\mathrm{erg~s^{-1}}$ \citep{moriya2013}. Thus, the interaction luminosity can have a comparable luminosity contribution to the recombination luminosity discussed so far.

The event rate of ILRTs is estimated to be $1-10$\% of core-collapse SNe \citep{cai2021,karambelkar2023}. If a fraction of ILRTs as well as SN~1987A-like SNe originate from BSGs, up to around 10\% of core-collapse SNe may originate from BSG explosions. We note again that some ILRTs are known to originate from low-mass progenitors (e.g., SN~2008S) and not all of ILRTs originate from BSGs.

The second group of light curves consist of various SN~1987A-like SNe, that result from high \Mni explosions of the BSG models (Fig.~\ref{fig:comparison}, bottom panel). These peculiar Type~II SNe are similar in their light-curve shapes to the prototype SN~1987A, and can be well-explained by high-energy explosions ($E\approx1$\,B) with $\Mni\geq0.07$~\Msun. 
A more precise reproduction of the observed light curves (such as in \cite{menon2019}), can be done by exploring the explosions of other BSG progenitor models.

\section{Summary}\label{sec:summary}
We explored explosion properties of BSGs especially when the amount of \Ni produced during the explosions is small. When $\Mni\lesssim 0.01~\Msun$, the characteristic slow rise in light curves observed in SN~1987A-like SN light curves may disappear, and BSG explosions are observed as low-luminosity (around $10^{40}-10^{41}~\mathrm{erg s^{-1}}$), short-plateau ($50-90$~days) SNe. This is because of the compactness of the BSG progenitors that allows the adiabatic cooling efficient. The exact threshold \Ni mass for BSG explosions to be observed as such Type~II SNe depends on the radii of the BSG progenitors. Low-energy, low-\Ni mass explosions of BSGs can result in transients similar to those of some ILRTs. Therefore, we conclude that a fraction of ILRTs may originate from BSG explosions. The upcoming Legacy Survey of Space and Time (LSST) by the Vera C. Rubin Observatory would allow us to explore the nature of these faint transients and they may be able to probe diverse deaths of BSGs.

\begin{ack}
We thank Yongzhi Cai for sharing the photometric data of ILRTs, along with Norbert Langer and Alexander Heger for vital discussions that have shaped this paper. TJM is supported by the Grants-in-Aid for Scientific Research of the Japan Society for the Promotion of Science (JP24K00682, JP24H01824, JP21H04997, JP24H00002, JP24H00027, JP24K00668) and by the Australian Research Council (ARC) through the ARC's Discovery Projects funding scheme (project DP240101786). AM acknowledges the support of the Juan de la Cierva-Incorporación in conducting this research. AM also thanks the NAOJ Visiting Joint Research program for hosting her as an invited researcher. Numerical computations were in part carried out on PC cluster at Center for Computational Astrophysics, National Astronomical Observatory of Japan. This work was supported by the NAOJ Research Coordination Committee, NINS (NAOJ-RCC-2401-0402).
\end{ack}

%\appendix 

%%%
% See the manual for the detail.
%%%
\bibliographystyle{apj}
\bibliography{pasj}

\end{document}